\documentclass[%
 reprint,
superscriptaddress,
 amsmath,amssymb,
 aps,
]{revtex4-1}

\usepackage{graphicx}
\usepackage{dcolumn}
\usepackage{bm}
\usepackage{mathtools}
\usepackage{gensymb}
\usepackage{amsmath}
\usepackage{qcircuit}
\DeclarePairedDelimiter\bra{\langle}{\rvert}
\DeclarePairedDelimiter\ket{\lvert}{\rangle}

\usepackage{xr}
\makeatletter
\newcommand*{\addFileDependency}[1]{
  \typeout{(#1)}
  \@addtofilelist{#1}
  \IfFileExists{#1}{}{\typeout{No file #1.}}
}
\makeatother

\newcommand*{\myexternaldocument}[1]{%
    \externaldocument{#1}%
    \addFileDependency{#1.tex}%
    \addFileDependency{#1.aux}%
}

\myexternaldocument{supplement}

\begin{document}


\title{Quantum Refrigeration with Indefinite Causal Order}

\author{David Felce}
\email{david.felce@physics.ox.ac.uk}
\affiliation{Clarendon Laboratory, Department of Physics, University of Oxford, England}

\author{Vlatko Vedral}
\affiliation{Clarendon Laboratory, Department of Physics, University of Oxford, England}
\affiliation{Centre for Quantum Technologies, National University of Singapore, Block S15, 3 Science Drive 2, Singapore}
\affiliation{Department of Physics, National University of Singapore, Science Drive 3, Blk S12, Level 2, Singapore 1175512}

\begin{abstract}
We propose a thermodynamic refrigeration cycle which uses Indefinite Causal Orders to achieve non-classical cooling. The cycle cools a cold reservoir while consuming purity in a control qubit. We first show that the application to an input state of two identical thermalizing channels of temperature $T$ in an indefinite causal order can result in an output state with a temperature not equal to $T$. We investigate the properties of the refrigeration cycle and show that thermodynamically, the result is compatible with unitary quantum mechanics in the circuit model but could not be achieved classically. We believe that this cycle could be implemented experimentally using tabletop photonics. Our result suggests the development of a new class of thermodynamic resource theories in which operations are allowed to be performed in an Indefinite Causal Order.
\end{abstract}

\maketitle


\paragraph*{Introduction---}
An indefinite casual order arises when the order in which events take place, or operations are performed, is in a quantum superposition. This enables higher order quantum operations that cannot be represented exactly using a standard quantum circuit \citep{PhysRevA.88.022318}. It has been proposed that a theory of quantum gravity might allow for the superposition of different spacetimes in such a way that the path of a quantum system is in a superposition - thus the system can interact with other systems in a superposition of causal orders \citep{Hardy_2007}. \par
A recent framework \citep{PhysRevA.88.022318} in which indefinite casual orders can be considered, the quantum SWITCH, has generated a good deal of discussion. It has been shown that the utilisation of indefinite causal orders can yield advantages in quantum computation \citep{Colnaghi2012,PhysRevLett.113.250402},  communication \citep{PhysRevLett.120.120502,2018arXiv180906655S,chiribella2018indefinite,2019arXiv190201807P}, and metrology \citep{2018arXiv181207508M,Zhao:19}. Recent experiments \citep{Procopio2015,PhysRevLett.121.090503,goswami2018communicating,PhysRevLett.122.120504,PhysRevLett.124.030502} have even been claimed to demonstrate the use of indefinite causal orders. \par
Encouraged by results \citep{PhysRevLett.120.120502,2018arXiv180906655S} considering Indefinite Causal Orders in Quantum Information, we ask a question in the closely related field of Quantum Thermodynamics - namely, does a quantum mechanical uncertainty in the causal structure underlying a system offer an advantage in performing thermodynamic tasks? We will answer this question in the affirmative. \par
We start by calculating the output of thermalizing channels in an indefinite causal order (ICO). We then consider a thermodynamically equivalent circuit picture which produces the same output. Finally, using our first result we construct a thermodynamic cycle utilizing indefinite causal orders which transfers heat from cold to hot reservoirs. \par
In this paper we represent thermodynamic operations as quantum channels. These are completely positive trace preserving (CPTP) maps, which describe general transformations of the density operators of a system which is embedded in a larger system obeying unitary dynamics. A quantum channel $\mathcal{N}$ acting on a state $\rho$ admits the Kraus decomposition $\mathcal{N}(\rho) = \sum_i K_i \rho K^\dagger_i$, where the operators $\{K_i\}$ satisfy $\sum_i K^\dagger_i K_i = I$. \par
\begin{figure}[ht]
 \includegraphics[width=\linewidth]{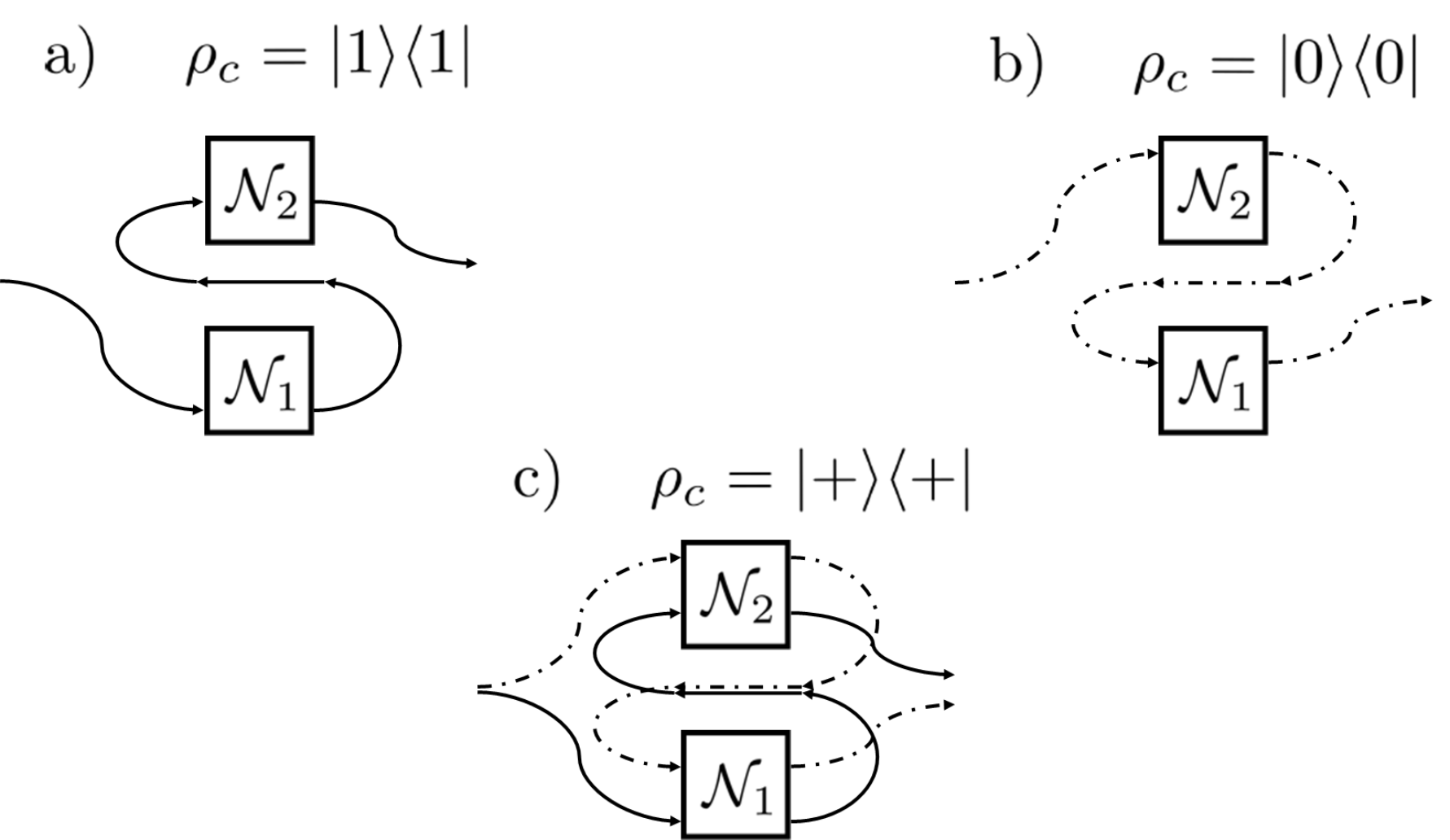}
 \caption{(a) and (b) illustrate channels $\mathcal{N}_1$ and $\mathcal{N}_2$ placed in a definite order, corresponding to the control qubit being in state $\ket{1}\bra{1}$ and $\ket{0}\bra{0}$ respectively. In (c) the quantum SWITCH places the channels in a superposition of causal orders. It entangles the order of the two channels with the state of the control qubit, in this case $\ket{+}\bra{+}$.}
 \label{fig:QSWITCH}
\end{figure}

Ordinarily, operations, represented here by quantum channels, are applied in a fixed causal order. However, it has been shown that applying channels in an indefinite causal order yields interesting results. One paradigm in which indefinite causal order is implemented is the quantum SWITCH. The quantum SWITCH, illustrated in Fig. \ref{fig:QSWITCH}, takes a pair of channels and applies them in an order which is correlated with the state of a control qubit. \par
\indent

The Kraus operators of the map resulting from the quantum SWITCH of channels $N_1$ and $N_2$ are

\begin{equation} 
W_{ij} = \ket{0}\bra{0}_c \otimes K^{(2)}_iK^{(1)}_j + \ket{1}\bra{1}_c \otimes K^{(1)}_jK^{(2)}_i,
\label{eq:Wij}
\end{equation}

where the subscript $c$ denotes the control qubit and the operators $K^{(2)}_i$ and $K^{(1)}_j$ denote the Kraus operators for $\mathcal{N}_2$ and $\mathcal{N}_1$ respectively. These Kraus operators act on a target quantum state $\rho$ and a control state $\rho_c$ so that the quantum SWITCH of the two channels gives:

\begin{equation} 
S(\mathcal{N}_1,\mathcal{N}_2)(\rho_c\otimes\rho) = \sum_{i,j}W_{ij}(\rho\otimes\rho_c)W^\dagger_{ij}.
\label{eq:SN1N2}
\end{equation}

\paragraph*{Thermalizing channels in an Indefinite Causal Order---}\label{sec:SIS}

Recent work \citep{PhysRevLett.120.120502,2018arXiv180906655S} has demonstrated the counter-intuitive result that two completely depolarizing channels, when placed in an indefinite causal order, can transmit information. This is despite the fact that these channels have no information capacity when used in any fixed or classically random ordering. The action of the fully depolarizing channel $\mathcal{N}^D$ can be represented by uniform randomization over $d^2$ orthogonal unitary operators $U_i$ 

\begin{equation} 
\mathcal{N}^D(\rho) = \textrm{Tr}[\rho]\frac{I}{d} = \frac{1}{d^2}\sum^{d^2}_{i}U_{i}\rho U^\dagger_{i}.
\label{eq:depol}
\end{equation}

It is easy to see that this channel is constant and as a result it has no information capacity. However, when two of these channels are used in an indefinite causal order, non-zero Holevo information can be transmitted. We would like to exploit the intimate relationship between information theory and thermodynamics, so we turn to thermalizing channels - the closest equivalent to depolarizing channels in the world of thermodynamics. A thermalizing channel is defined by a certain inverse temperature $\beta$ and maps any input state $\rho$ to the output state $T$ which is a thermal state with effective inverse temperature $\beta$. The density operator representing the state $T$ in the energy basis is a diagonal matrix with entries $\frac{1}{Z}e^{-\beta E_j}$ where $E_j$ are the energy eigenvalues and $Z$ is a normalizing partition function. The action of the thermalizing channel $\mathcal{N}^T$ is

\begin{equation} 
\mathcal{N}^T(\rho) = \textrm{Tr}[\rho] T = \frac{1}{d}A\left(\sum^{d^2}_{i}U_{i}\rho U^\dagger_{i}\right) A^\dagger,
\label{eq:therm}
\end{equation}

where $A$ is the square root of the diagonal matrix $T$. It can be easily seen from the second equality in equations (\ref{eq:depol}) and (\ref{eq:therm}) that from the depolarizing channel to the thermalizing channel the Kraus operators change from

\begin{equation} 
K^D_i = \frac{1}{d}U_i \longrightarrow K^T_i = \sqrt{\frac{1}{d}}AU_i.
\nonumber
\label{eq:kraus}
\end{equation}

The new operators do indeed satisfy the requirement for Kraus operators that $\sum_iK^\dagger_i K_i = I$, since the set $\{U^\dagger_i\}$ is also a set of orthogonal unitary operators and so equation (\ref{eq:depol}) can be used to obtain identity. \par
We now calculate the result when two identical thermalizing channels in an indefinite causal order are applied to a state $\rho$ and control system $\rho_c$. If the control system is initialized in the state $\rho_c = \ket{\psi_c}\bra{\psi_c}$ where $\ket{\psi_c} = \sqrt{\alpha}\ket{0} + \sqrt{1-\alpha}\ket{1}$, then the output state after interaction with the thermalizing channels in an indefinite causal order will be:

\begin{widetext}

\begin{align}
  \phantom{S(\mathcal{N}^T,\mathcal{N}^T)(\rho_c\otimes\rho)}
  &\begin{aligned}
    \mathllap{S(\mathcal{N}^T,\mathcal{N}^T)(\rho_c\otimes\rho)} &= \frac{1}{d^2}\sum_{i,j}\Big( \alpha\ket{0}\bra{0}_c\otimes AU_iAU_j\rho U^\dagger_jA^\dagger U^\dagger_iA^\dagger + (1-\alpha)\ket{1}\bra{1}_c\otimes AU_jAU_i\rho U^\dagger_iA^\dagger U^\dagger_jA^\dagger\\
      &\quad +\sqrt{\alpha(1-\alpha)}\ket{0}\bra{1}_c\otimes AU_iAU_j\rho U^\dagger_iA^\dagger U^\dagger_jA^\dagger +\sqrt{\alpha(1-\alpha)}\ket{1}\bra{0}_c\otimes AU_jAU_i\rho U^\dagger_jA^\dagger U^\dagger_iA^\dagger \Big) \nonumber
  \end{aligned}\\
  &\begin{aligned}
    &= \sum_{i}\frac{\alpha}{d}\ket{0}\bra{0}_c\otimes AU_iTU^\dagger_iA^\dagger + \sum_{j}\frac{1-\alpha}{d}\ket{1}\bra{1}_c\otimes AU_jTU^\dagger_jA^\dagger\\
      &\quad + \frac{\sqrt{\alpha(1-\alpha)}}{d}\sum_j\left( \ket{0}\bra{1}_c\otimes T\textrm{Tr}[U_j\rho A] U^\dagger_jA^\dagger + \ket{1}\bra{0}_c\otimes A\textrm{Tr}[A^\dagger\rho U^\dagger_j]U_jT \right) \nonumber
  \end{aligned}\\
  &\begin{aligned}
    &= \left(\alpha\ket{0}\bra{0}_c+(1-\alpha)\ket{1}\bra{1}_c\right)\otimes T + \sqrt{\alpha(1-\alpha)}\left(\ket{0}\bra{1}_c + \ket{1}\bra{0}_c\right)\otimes T\rho T. \label{eq:ICOtherm}
  \end{aligned}
\end{align}

\end{widetext}

The first equality is given by the application of the quantum SWITCH in equation (\ref{eq:SN1N2}). The second equality is the action of the depolarizing channel in equation (\ref{eq:depol}). In the third equality, the diagonal terms are computed by another application of the depolarizing channel, and the off diagonal terms are computed using the fact that the operators $U_j$ form an orthonormal basis for the set of $d$ x $d$ matrices. \par
Instead of thermalizing to the expected thermal state $T$, the final state of the system is in general different. We note that the final state of the system is entangled with the state of the control qubit. If the control qubit is measured in the computational basis, then the system will be found in the thermal state $T$. However if the control qubit is measured in the $\{\ket{+}_c,\ket{-}_c\}$ basis then we recover:

\begin{equation} 
_c\bra{\pm}S(\mathcal{N}^T,\mathcal{N}^T)(\rho_c\otimes\rho)\ket{\pm}_c = \frac{T}{2} \pm \sqrt{\alpha(1-\alpha)} T\rho T,
\label{eq:CondICO}
\end{equation}

with,

\begin{equation} 
p_{\pm} = \textrm{Tr}\Big[\frac{T}{2} \pm \sqrt{\alpha(1-\alpha)} T\rho T\Big]
\label{eq:MeasureProbs}
\end{equation}

giving the probabilities of measuring the control qubit in the states $\ket{+}_c$ or $\ket{-}_c$. \par
Two features of this result should be noted. First, the final state preserves some dependence on the initial state $\rho$. This is counter-intuitive, since ordinarily the thermalizing channel would remove any such dependence. Secondly, the effective temperature of the final state will be different than that of the expected thermal state $T$. The temperature of the final state will be higher or lower, depending on whether the $\ket{+}_c$ or $\ket{-}_c$ state is measured at the control qubit. It is this fact that we will exploit to perform refrigeration.

\paragraph*{Equivalent Circuit Picture---}

It is possible to construct a Unitary quantum circuit which acts on the system plus environments and has the same output as the Quantum Switch of the thermalizing channels.

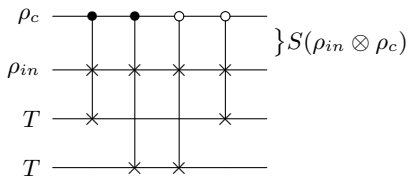
\begin{figure}[ht]
\begin{align} 
\Qcircuit @C=1.5em @R=2em {
\lstick{\rho_c} & \ctrl{2} & \ctrl{3} & \ctrlo{3} & \ctrlo{2} &  \qw & & \dstick{\big\}S(\rho_{in} \otimes\rho_c)} \\
\lstick{\rho_{in}} & \qswap & \qswap & \qswap & \qswap &  \qw \\
\lstick{T} & \qswap & \qw & \qw & \qswap &  \qw \\
\lstick{T} & \qw & \qswap & \qswap & \qw &  \qw 
}
\nonumber
\end{align}
 \caption{A circuit with an output equivalent to that of the quantum SWITCH of thermalizing channels.}
\end{figure}

Here the order in which the swaps occur is determined by the state of the control qubit. After this circuit the marginal state on the upper two wires is identical to $S(\mathcal{N}^T,\mathcal{N}^T)(\rho\otimes\rho_c)$ in equation (\ref{eq:ICOtherm}). Although there are in principle an infinite number of circuits, with different environments, which give this output, this particular representation is helpful for thermodynamic considerations, because the environments (two bottom inputs $T$) can be thought of as qubits randomly drawn from a reservoir which is a thermal bath of qubits each with thermal state $T$. For any unitary circuit, the free energy of the output is equal to the free energy of the inputs. The only subsystem of the input state that has non-zero free energy in the case when $\rho_{in} = T$ is the control, since all the other inputs are in a thermal state. In this case, which we discuss below, the free energy in the control is then distributed between the control and the system. Although the circuit is thermodynamically equivalent to the quantum SWITCH of two thermalizing channels, it uses each channel twice, and so is not equivalent in every sense. 

\paragraph*{ICO refrigerator---}

A natural question is whether the ability to perform operations in an ICO can be useful in performing thermodynamic tasks. The fact that a system can `thermalize' with identical reservoirs and the resulting state have a temperature not equal to the temperature of the reservoirs suggests that this is the case. \par
For simplicity, we will consider the case where the system $\rho$ has two energy eigenstates - a ground and an excited state, $\ket{g}$ and $\ket{e}$. This assumption simplifies our considerations because in a two-level system every state can be assigned a consistent effective temperature. The Hamiltonian for the system is defined as $H = \Delta \ket{e}\bra{e}$. Since we are interested in these operations in the context of thermodynamic work cycles, let us restrict ourselves to thermal initial states. Then the density matrix for the initial state is

\begin{equation} 
\rho = \frac{1}{Z_\rho}\begin{pmatrix} 
1 & 0 \\
0 & e^{-\beta_\rho \Delta} 
\end{pmatrix},
\quad
Z_\rho = 1+e^{-\beta_\rho \Delta},
\nonumber
\end{equation}

where $\beta_\rho$ is the state's inverse temperature. We also note that the state obtained after thermalizing classically with the reservoir is

\begin{equation} 
T = \frac{1}{Z_T}\begin{pmatrix} 
1 & 0 \\
0 & e^{-\beta_T \Delta} 
\end{pmatrix},
\quad
Z_T = 1+e^{-\beta_T \Delta}.
\nonumber
\end{equation}

Let us take our initial state and the reservoirs to be at the same temperature, $i.e$ $\beta_\rho = \beta_T = \beta$, and $Z_\rho = Z_T = Z$. Using equation (\ref{eq:CondICO}), and setting $\alpha = \frac{1}{2}$ (an equal superposition of orders), the resulting state after ICO thermalization and control system measurement is:

\begin{equation} 
\rho_{ICO} = \frac{\rho_{ICO}'}{\textrm{Tr}[\rho_{ICO}']},\quad
\rho_{ICO}' = \begin{pmatrix} 
1 \pm \dfrac{1}{Z^2} & 0 \\[2ex]
0 & e^{-\beta \Delta} \pm \dfrac{e^{-3\beta \Delta}}{Z^2} 
\end{pmatrix}.
\nonumber
\end{equation}

After this transformation, the system is at a different temperature than the reservoir. Hence one can create a thermodynamic cycle, the operation of which transfers heat from an ensemble of cold reservoirs to a hotter reservoir (inverse temperatures $\beta_C$ and $\beta_H$ respectively), while consuming purity in the control qubit, or, equivalently, consuming work due to the erasure of the results of measurement. \par
The working system starts at the temperature of the cold reservoirs. 

\begin{figure}[ht]
 \includegraphics[width=\linewidth]{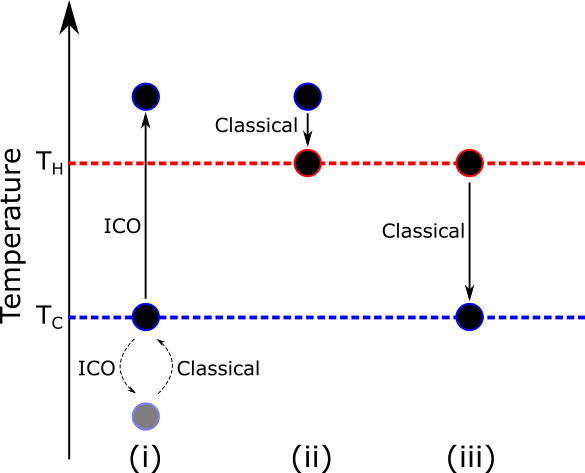}
 \caption{The three steps of the refrigeration cycle of the ICO Fridge. The black dot represents the working system, and the colour of the outline indicates the temperature of the last reservoir(s) with which it has interacted. The dotted lines in step (i) represents the operation in the event of a measurement of $\ket{+}_c$ (the undesired outcome) for the state of the control system.}
 \label{fig:Cycle}
\end{figure}

In the first step (i), two of the cold reservoirs are placed in an indefinite causal order using the quantum SWITCH, and the working system `thermalizes' with them. The control qubit is measured. If the control qubit is found in the state $\ket{+}_c$ (in this case heat will have passed from the system into the reservoirs, which is the reverse of what is desired) then the system is thermalized classically with another cold reservoir (indicated by the dotted lines in the figure), which reverses the heat flow, returning to the initial state with the only side effect being that a measurement has been made. Then the step is repeated until $\ket{-}_c$ is measured. \par
In the second step (ii), the system is thermalized classically with the hot reservoir. In the third step (iii), the system is thermalized classically with a cold reservoir, and all the cold reservoirs are thermalized classically with one another, so that they remain identical. \par
In step (i), heat is transferred from the cold reservoirs into the working system. In step (ii), heat is transferred from the working system to the hot reservoir. In step (iii) heat is transferred from the working system to the cold reservoirs, but this heat change is smaller in magnitude than that in step (i), so overall, heat is transferred out of the cold reservoir. \par
It is important to note that despite the fact that a measurement is made in the process of the operation of the refrigeration cycle, and the next step is conditioned on the result of this measurement, the cycle does not rely on post-selection to gain its advantage. The outcomes where an undesired result is measured are fully accounted for in the consideration of the average thermodynamic performance of the cycle. \par
In fact, there is an equivalent cycle without a projective measurement at all. Instead of measuring the control qubit at the end of step (i), the execution of step (ii) is controlled (on the state $\ket{-}_c$) by the control qubit. and step (iii) is carried out as normal. The end result will be a superposition of a branch where the control qubit is in the state $\ket{+}_c$ and no heat has been transferred, and a branch where the control qubit is in the state $\ket{-}_c$ and heat has been transferred from the cold to the hot reservoirs. To run the cycle again, the old control qubit is discarded and a new control qubit in the state $\ket{+}_c$ must be introduced. When implemented in this way, the cycle consumes purity in the control qubit, rather than the work cost of measurement, to run. The two implementations are thermodynamically equivalent.

\paragraph*{Performance of the refrigerator---}

The performance of the ICO fridge is dependent on the temperatures of the hot and cold reservoirs and the Hamiltonian of the working substance. In order to calculate heat flows, we assume that all thermalizations are isochoric, $i.e.$ that all changes in internal energy of the working system correspond to heat flows to or from the reservoirs. \par
It is then possible to calculate the heat change per cycle of the cold reservoir, which is given by

\begin{equation} 
\delta Q_{Cold}^{cycle} = \Big[\dfrac{e^{-\beta_H\Delta}}{1+e^{-\beta_H\Delta}}-\dfrac{2e^{-\beta_C\Delta}+1}{3(1+e^{-\beta_C\Delta})}\Big]\Delta.
\label{eq:dS}
\end{equation}

It can be seen that a `positive refrigeration condition' which must be satisfied if the cycle is to transfer heat from the cold reservoirs to the hot reservoir, instead of from hot to cold, is given by

\begin{equation} 
\frac{2e^{-\beta_C\Delta} + 1}{3\left(e^{-\beta_C\Delta} + 1\right)} > \frac{e^{-\beta_H\Delta}}{1+e^{-\beta_H\Delta}}.
\label{eq:PRC}
\end{equation}

This condition is equivalent to ensuring that after step (i) of the cycle, the effective temperature of the working substance is higher than the temperature of the hot reservoir. See Appendix A for a derivation of equations (\ref{eq:dS}) and (\ref{eq:PRC}). \par
It is possible, by allowing $\beta_H$ to vary, to find the maximum possible heat extraction per cycle for a given $\beta_C$. Unfortunately a closed form solution is impossible since we obtain a transcendental equation. \par
Despite this, it is clear that in the limit where both reservoir temperatures are low, i.e.

\begin{equation} 
\beta_C\Delta > \beta_H\Delta >> 1,
\nonumber
\label{eq:LowT}
\end{equation}

the positive refrigeration condition (\ref{eq:PRC}) is satisfied. In this case we have the approximate relation

\begin{equation} 
\delta Q_{Cold}^{total} = -\frac{1}{3}\Delta.
\label{eq:dSapprox}
\end{equation}

This is a striking result as it does not depend strongly on the temperature of either reservoir. However, in these regimes, the probability of measuring $\ket{-}_c$ becomes exponentially small, so step (i) of the cycle must be repeated many times. \par
The cost of the operation of the fridge is that a measurement must be made, and thus work done, every time the cycle is run (see Appendix B). The coefficient of performance can be calculated by dividing the heat transfer from the cold reservoir by the work cost of the measurements \citep{abdelkhalek2016fundamental}, which comes from Landauer's erasure \citep{Landauer1961}. The efficiency of the fridge approaches the Carnot efficiency for a very low temperature cold reservoir and a suitably chosen hot reservoir, but for most parameters the efficiency is much lower. The reason for the cycle's reduced efficiency is that during step (i) there are correlations produced in the reservoirs which are not utilised, but simply allowed to `thermalize away'. In addition, the thermalizing procedures in steps (ii) and (iii) are irreversible isochoric thermalizations. It is likely that a related scheme which instead used reversible processes would achieve an efficiency closer to the Carnot bound for more general parameters.

\paragraph*{Discussion---}

Our results are of interest for resource theories of thermodynamics, since they show that a certain thermodynamic task - the transfer of heat from a cold to a hot reservoir - can be achieved by using thermalizing channels alone, if one is allowed to apply them in an indefinite causal order (as well as measure a control qubit). This task would clearly be impossible if the channels could only be applied in a fixed or classically random order. This suggests the development of a new class of thermodynamic resource theories, in which operations can be performed in indefinite causal orders. We expect that these theories might have exotic properties that differ substantially from their counterparts with fixed casual orders.\par
Our results demonstrate that in at least one case, a quantum uncertainty in the causal structure of a thermodynamic processes can provide a thermodynamic advantage. It is natural to believe that there might be other thermodynamic protocols that can be made possible or enhanced with the use of indefinite causal order. We hope that this work will encourage more investigation into new avenues utilizing indefinite casual orders in thermodynamics. \par
We believe that an experimental realisation of refrigeration using indefinite causal orders is achievable. Tabletop photonics experiments \citep{goswami2018communicating,PhysRevLett.122.120504,PhysRevLett.124.030502} have already been carried out to demonstrate enhanced communication using the quantum SWITCH. The step from these experiments to implementing the ICO refridgerator should be, in principle, not too great. An implementation of our equivalent circuit should also be achievable on many other qubit platforms using current technology. \par
We have also shown that the application of thermalizing channels in an indefinite causal order has a thermodynamically equivalent quantum circuit representation. However because the circuit uses each channel twice, it is not fully equivalent in every way.  Our results do not place a bound on the energy cost to create an indefinite causal order, since the work expended in the measurement of the control qubit is enough to account for the cooling of the cold reservoir. \par
The application of the pseudo-density matrix formalism \citep{Fitzsimons2015} to the two level system before and after its ICO interactions could provide an interesting insight into the causal structure underlying the operation of the quantum SWITCH. We leave this for future work.

\paragraph*{Acknowledgements---}

We would especially like to thank Daniel Ebler, Sina Salek, and Mile Gu for helpful discussions and comments, and Chiara Marletto, Felix Tennie, Benjamin Yadin, Samuel Kuypers, and Adrian Menssen for their constructive input. DF is supported by the EPSRC (UK) and by M squared. VV thanks the National Research Foundation, Prime Minister’s Office, Singapore, under its Competitive Research Programme (CRP Award No. NRF- CRP14-2014-02) and administered by Centre for Quantum Technologies, National University of Singapore.

\bibliography{references}

\end{document}


\begin{appendix}

\section{Derivation of Fridge Performance}

Here follows a derivation of equation (\ref{eq:dS}).

We take the initial temperatures of the hot and cold reservoirs to be $\beta_H$ and $\beta_C$ respectively. The working bit is initially in the state

\begin{equation} 
\rho_0 = \frac{1}{Z_C}\begin{pmatrix} 
1 & 0 \\
0 & r_C 
\end{pmatrix},
\quad
Z_C = 1+r_C ,
\nonumber
\end{equation}

where we have taken $r_x = e^{\beta_x \Delta}$ to ease the burden of notation.
Following an application of two cold reservoirs in an ICO, and the measurement of $\ket{-}_c$ as desired, in step (i) of the fridge operation, equation (\ref{eq:CondICO}), setting $\alpha = \frac{1}{2}$ gives the output state as

\begin{equation}
\rho_1' = \begin{pmatrix} 
\dfrac{1}{Z_C} & 0 \\
0 & \dfrac{r_C}{Z_C} 
\end{pmatrix} - \begin{pmatrix}
\dfrac{1}{Z_C^3} & 0 \\
0 & \dfrac{r_C^3}{Z_C^3} 
\end{pmatrix},
\nonumber
\end{equation}

where the prime denotes that normalization is required.
This can be rewritten as 

\begin{equation}
\rho_1' = \frac{1}{Z_C}\begin{pmatrix} 
1-\dfrac{1}{(1+r_C)^2} & 0 \\
0 & r_C-\dfrac{r_C^3}{(1+r_C)^2} 
\end{pmatrix},
\nonumber
\end{equation}

which simplifies for $r_c \neq 0$, giving the normalized state

\begin{equation}
\rho_1 = \frac{r_C+2}{3r_C+3}\begin{pmatrix} 
1 & 0 \\
0 & \dfrac{2r_C+1}{r_C+2} 
\end{pmatrix}.
\nonumber
\end{equation}

We can thus calculate the heat flow into the cold reservoir in step (i):

\begin{equation}
Q_{Cold}^{(i)} = -[\rho_1^{(2,2)}-\rho_0^{(2,2)}]\Delta = \dfrac{r_C - 1}{3(1+r_C)}\Delta.
\nonumber
\end{equation}

Next, in step (ii). the system thermalizes classically with the hot reservoir, after which

\begin{equation}
    \rho_2 = \dfrac{1}{1+r_H}
    \begin{pmatrix} 
        1 & 0 \\
        0 & r_H 
    \end{pmatrix},
\nonumber
\end{equation}

with a corresponding heat transfer to the hot reservoir of

\begin{equation}
Q_{Hot}^{(ii)} = -[\rho_2^{(2,2)}-\rho_1^{(2,2)}]\Delta = \Big[\dfrac{2r_C + 1}{3(1+r_C)}-\dfrac{r_H}{1+r_H}\Big]\Delta.
\nonumber
\end{equation}

Next, in step (iii), the system is thermalized classically with one of the cold reservoirs, after which

\begin{equation}
    \rho_3 = \dfrac{1}{1+r_C}
    \begin{pmatrix} 
        1 & 0 \\
        0 & r_C 
    \end{pmatrix},
\nonumber
\end{equation}

with a corresponding heat transfer to the cold reservoirs of

\begin{equation}
Q_{Cold}^{(iii)} = -[\rho_3^{(2,2)}-\rho_2^{(2,2)}]\Delta = \Big[\dfrac{r_H}{1+r_H}-\dfrac{r_C}{1+r_C}\Big]\Delta.
\nonumber
\end{equation}

The total heat change in the Cold reservoir is thus

\begin{equation}
Q_{Cold}^{cycle} = Q_{Cold}^{(i)} + Q_{Cold}^{(iii)} = \Big[\dfrac{r_H}{1+r_H}-\dfrac{2r_C+1}{3(1+r_C)}\Big]\Delta.
\nonumber
\end{equation}

\section{Work Cost and Coefficient of Performance}

The work cost associated with the running of the cycle arises from the measurement performed in step (i). Each time a measurement is made, the result of the measurement must be stored in a register. The cycle then proceeds according to the value in the register. Once the result of the measurement is no longer needed then the register must be erased in order to return it to its initial state. Since the measurement itself can be performed reversibly and with no energy cost (since the control qubit is assumed to have a zero Hamiltonian), the energy cost comes from this process of erasure. \par
In the case of a projective measurement, with operators $\{P_k\}$, on a system $\rho$, the probability of a given outcome is given by $p_k = \textrm{Tr}[P_k\rho]$. After the measurement the Shannon entropy of the register $\rho_M$ is therefore

\begin{equation}
    S(\rho_M) = -\sum_k p_k \textrm{ln}(p_k).
    \nonumber
\end{equation}

The energy cost to reset the register in contact with a resetting reservoir with inverse temperature $\beta_R$ is then

\begin{equation}
    \Delta W_E = \frac{1}{\beta_R}S(\rho_M).
    \nonumber
\end{equation}

In the case of the measurement in step (i) of our cycle, the probabilities for measurement are given by equation (\ref{eq:MeasureProbs})

\begin{equation} 
p_{\pm} = \textrm{Tr}\Big[\frac{T}{2} \pm \sqrt{\alpha(1-\alpha)} T\rho T\Big],
\nonumber
\end{equation}

and so the work cost for erasure is

\begin{equation}
    \Delta W_E = \frac{1}{\beta_R}\sum_\pm p_\pm \textrm{ln}(p_\pm).
    \nonumber
\end{equation}

In the operation of the cycle, step (i) must be repeated until $\ket{-}_c$ is measured. Therefore on average the number of measurements performed per cycle will be

\begin{equation}
    \bar{n}_M = \frac{1}{p_-},
    \nonumber
\end{equation}

with a corresponding average work cost per cycle of

\begin{equation}
    \Delta W^{cycle} = \frac{1}{\beta_R p_-}\sum_\pm p_\pm \textrm{ln}(p_\pm).
    \nonumber
\end{equation}.

The coefficient of performance can then be calculated by dividing the heat transfer from the cold reservoir in equation (\ref{eq:dS}) by this work cost.

\begin{equation}
    \eta_{COP} = -\frac{Q_{Cold}^{cycle}}{\Delta W^{cycle}}.
    \nonumber
\end{equation}

\end{appendix}